\documentclass[conference]{IEEEtran}

\usepackage{ifpdf}
\usepackage{cite}
\usepackage{algorithmic}
\usepackage{algorithm}
\usepackage{color}
\ifCLASSINFOpdf
   \usepackage[pdftex]{graphicx}

\else
 \usepackage[dvips]{graphicx}
\fi
\usepackage[cmex10]{amsmath}
\usepackage{epstopdf}
\usepackage{array}
\usepackage{balance}
\usepackage[top = 0.75 in, bottom = 0.95in, left = 0.625in, right = 0.625in]{geometry}
\usepackage{eqparbox}

\usepackage[tight,footnotesize]{subfigure}

\usepackage{fixltx2e}

\usepackage{float}

\usepackage{dblfloatfix}
\usepackage{url}
\usepackage{cite}
\usepackage{amssymb}
\begin{document}

\title{Relay--Reconfigurable Intelligent Surface Cooperation for Energy-Efficient Multiuser Systems}

\setlength{\columnsep}{0.21 in}

\author{\IEEEauthorblockN{Mohanad Obeed and Anas Chaaban}
\IEEEauthorblockA{School of Engineering, University of British Columbia, Kelowna, Canada.\\ 
Email: \{mohanad.obeed, anas.chaaban\}@ubc.ca
}}

\maketitle

\begin{abstract}
Reconfigurable intelligent surfaces (RIS) have drawn considerable attention recently due to their controllable scattering elements that are able to direct electromagnetic waves into desirable directions. Although RISs share some similarities with relays, the two have fundamental differences impacting their performance. To harness the benefits of both relaying and RISs, a multi-user communication system is proposed in this paper wherein a relay and an RIS cooperate to improve performance in terms of energy efficiency. To utilize the RIS efficiently, the discrete phase shifts of the RIS elements are optimized along with the beamforming matrices at the transmitter and the relay, targeting the minimization of the total transmit power subject to a quality-of-service (QoS) constraint. Then, two suboptimal efficient solutions are proposed for the resulting discrete and non-convex problem, one based on singular value decomposition (SVD) and uplink-downlink duality and the other is based on SVD combined with zero-forcing. Simulations show that the proposed solutions outperform a system with either a relay or an RIS only, especially when both are closer to the users than to the base-station.
\end{abstract}
\IEEEpeerreviewmaketitle

\section{Introduction}
Wave propagation in wireless communication systems is affected by scattering, delays, reflections, and diffractions. Such factors make the received signal consist of multiple random, delayed, and attenuated copies of the transmitted signal, which impacts the achievable rate, energy efficiency, and coverage probability. Thus, adding some level of reconfigurability to the propagation medium can improve performance. Recently, a solution has been proposed to achieve this goal via deploying a reconfigurable intelligent surface (RIS), a two dimensional large surface consisting of digitally-controllable reflecting elements \cite{alouiniR26,cuiR22, WuR21}. 

The elements of an RIS can be controlled to strengthen the received signal or weaken the received interference. This is done by adjusting the phase shifts between the incident and reflected waves at each element so that the received waves add up constructively or distructively at a receiver. In the literature, several papers investigate how these phase shifts can be optimized along with the beamforming vectors at the BS, with the goal of improving performance, showing that RISs can play an important role in future communication networks. For instance, an RIS can significantly improve the sum-rate in a MIMO single user or multiuser system \cite{9110889}, increase the energy efficiency of the communication system \cite{WuR1, huang2019reconfigurable}, decrease inter-cell interference \cite{9090356}, enhance the secrecy rate \cite{8723525}, and extend the coverage of millimetre wave communications \cite{9226616}. Interestingly, as shown numerically in \cite{WuR14}, the SNR of the RIS-assisted system with $N=100$ reflecting elements is improved by 10 dB compared to the non RIS-assisted systems, and the SNR increases proportional to $N^2$ as $N$ increases. 

Although RISs can be used to improve the end-to-end channel quality, their main limitation is that the equivalent value of the reflected channel (base-station (BS) to RIS to user) is weak compared to the direct channel between the transmitter and the receiver, since this channel is the product of the BS-RIS channel and the RIS-user channel. This limitation can be overcome if the RIS is replaced by a relay node (both of which share the forwarding functionality), at the expense of additional energy consumption and complexity at the relay. The authors in \cite{HuangR3} compare RIS- and relay-assisted systems and show that the RIS-assisted system is able to provide up to 300\% higher energy efficiency compared to the use of a multi-antenna angular amplify-and-forward (AF) relaying system. This large gain is due to the fact that an RIS consumes very little power to forward the impinging signal contrary to a relay. However, the authors in \cite{sonR25} show that an RIS-assisted system cannot provide an SNR higher than that in a MIMO decode-and-forward (DF) relay-assisted system for any value of $N$, again at the cost of more power consumption. This is due to fact that the equivalent channel of the DF relay is much better than the reflected channel caused by RIS.  
Authors of \cite{RIS_Relay} discuss similarities and differences between RISs and relays, and argue that an RIS-assisted system outperforms a relay-assisted one in terms of data rates when the RIS is sufficiently large. 

From the above, we can conclude that the RIS-assisted system  performs better than the AF relay-assisted system, but worse than a DF relay-assisted one when the relay is equipped with a massive antennas. However, having the number of reflecting elements at the RIS much larger than the number of antennas at the DF relay, the RIS would perform better in terms of data rates or energy efficiency. 
In general, each of the DF relay and the RIS have their own pros and cons in terms of energy efficiency, hardware and software complexity, range, etc., and it is interesting to combine the two in a system to harness both gains from the RIS and the relay. 
Therefore, in this paper, we study a multi-user system employing both a DF relay and an RIS, and optimize their parameters (beamforming and reflection phases) to improve the multi-user system energy efficiency. 

We consider a system consisting of a BS, relay, RIS, and multiple users, and aim for minimizing the total transmit power (at both the BS and the relay) under quality-of-service (QoS) constraints represented in terms of achievable rates. We propose to place the RIS in the vicinity of the relay to enable better utilization of the reflecting elements in both hops (BS-to-relay and relay-to-users). We formulate the achievable rates, and then optimizes the beamforming matrices at the BS and relay and the discrete phase shifts at the RIS to minimize the total power subject to the required QoS. Since the problem is discrete and non-convex, we propose a suboptimal yet effective solution that uses an inner optimization to select the beamforming matrices of the BS and the relay, and an outer optimization to select the phase shifts for both hops. The inner optimization is based on singular value decomposition (SVD) and uplink-downlink duality and the other optimization is based on SVD combined with zero-forcing. Simulation results show the advantage of using both an RIS and a DF relay relative to using one of the two, in terms of energy efficiency.

We note that while all the aforementioned papers consider a system with an RIS or a relay, but not both, \cite{ying2020relay,9225707} study a system that consists of both an RIS and a relay. However, \cite{ying2020relay,9225707} focus on a single-user, single-antenna transmitter and receiver system and maximize the achievable rate, whereas our works considers a multi-antenna multi-user system which is more practical and challenging. 
Next, we present the system model and the achievable rates. We optimize the system in Sec. \ref{Sec:Form} and simulate its performance in Sec. \ref{Sec:Sim}. Finally, we conclude in Sec. \ref{Sec:Conc}.

\section{System Model}

\begin{figure}[!t]
\centering
\includegraphics[width=3.4in]{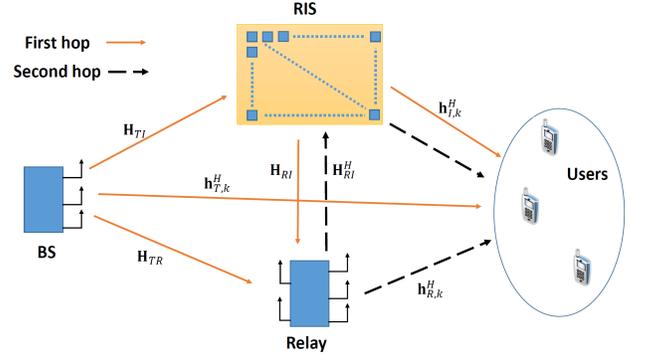}
\caption{System model consisting of a base-station, a relay, RIS, and $K$ users. The system is operated in a half-duplex mode, where the solid arrows indicate signal flow in the first phase, while dashed arrows indicate the signal flow in the second phase.}
\label{SM}
\end{figure}

We consider a system consisting of $K$ single-antenna users, a half-duplex DF relay with $N$ antennas, a BS with $M$ antennas, and an RIS with $L$ elements (Fig. \ref{SM}). The RIS is located near the relay to make it useful in both hops. Moreover, the RIS and relay are assumed to be closer to the users than to the BS.

We consider downlink transmission and assume that the system is operated in a half-duplex mode with two phases. In phase 1, the BS sends messages intended to users $1,\ldots,K$ encoded with rates $R_1,\ldots,R_K$ to the relay, and the relay decodes these messages. In phase 2, the relay forwards the messages to the users, and the users combine the received signals from both phases and decode their desired messages. Note that due to this, although the encoding rate is $R_k$, the overall rate achieved by user $k$ is $R_k/2$. The role of the RIS is to reflect the transmitted signal from the transmitter to the relay and the users in the first phase, and from the relay to the users in the second phase. The phase shift imposed by each reflecting element is assumed to be discrete, with $2^b$ possible phases where $b$ is the number of control bits. 

We denote the channel coefficient matrices from the BS to the relay, RIS, and user $k$ by $\mathbf{H}_{TR}\in\mathbb{C}^{N\times M}$, $\mathbf{H}_{TI}\in\mathbb{C}^{L\times M}$, and $\mathbf{h}_{T,k}^H\in\mathbb{C}^{1\times M}$, respectively, from the RIS to the relay as $\mathbf{H}_{IR}\in\mathbb{C}^{N\times L}$, and from the relay and RIS to user $k$ by $\mathbf{h}_{R,k}^H\in\mathbb{C}^{1\times N}$ and $\mathbf{h}_{I,k}^H\in\mathbb{C}^{1\times L}$, respectively. We assume that the BS-RIS and BS-relay channels have a line-of-sight (LoS) component which can be ensured by proper placement of the relay and RIS, while the channels to the users are non LoS. The channels maintain their value during a coherence interval and change independently between intervals (block fading). We also assume that the channel state information (CSI) of all links is known at the BS and the relay. 

The two phases of the transmission process, each of length $T$ symbols where $T$ is the codeword length, are explained next.

\subsection{Phase 1}
In the first phase, the BS transmits a precoded signal $\mathbf{x}(t)$ that can be expressed as follows
\begin{equation}
\mathbf{x}(t)=\sum_{k=1}^K \mathbf{w}_ks_k(t), \ t=1,\ldots,T,
\end{equation}
where $\mathbf{w}_k \in \mathbb{C}^{M\times 1}$ is the bemaforming vector assigned to user $k$ and $s_k(t)$ is a symbol of the length-$T$ codeword to be sent to user $k$. The symbols $s_k(t)$ have unit power, and the transmit power of the BS in this phase is $\frac{1}{2}\sum_{k=1}^K \|\mathbf{w}_k\|^2$, where the $\frac{1}{2}$ is due to the half-duplex operation since the BS is active half of the time.

Each user receives a superposition of the transmitted signal through the direct links and the reflected links via RIS, given by 
\begin{equation}
y_{k,1}(t)=(\mathbf{h}_{I,k}^H\mathbf{\Theta}_1\mathbf{H}_{TI}+\mathbf{h}_{T,k}^H)\sum_{i=1}^K\mathbf{w}_is_i(t)+n_k(t),
\end{equation}
$t=1,\ldots,T$, where $\mathbf{\Theta}_1=\text{diag}[e^{j\theta_{1,1}}, e^{j\theta_{1,2}},\ldots,e^{j\theta_{1,L}}]$, $\theta_{1,\ell}$ is the phase shift of the $\ell$th RIS element in the first phase, and $n_k(t)$ is additive white Gaussian noise (AWGN) with zero mean and variance $\sigma^2$, independent through $t$. The user will combine this signal with the received signal in the second phase before decoding.

The received signal at the relay in this phase is given by
\begin{equation}
\mathbf{y}_{r}(t)= (\mathbf{H}_{TR}+\mathbf{H}_{IR}\mathbf{\Theta}_1\mathbf{H}_{TI})\sum_{i=1}^K\mathbf{w}_is_i(t)+\mathbf{n}_R(t),
\end{equation}
$t=1,\ldots,T$, where $\mathbf{n}_R(t)$ is the AWGN vector at the relay with zero mean and covariance matrix $\sigma^2\mathbf{I}$, independent through $t$.


The relay uses a decode-and-forward scheme. It decodes the users' codewords from $\mathbf{y}_{r}(t)$, $t=1,\ldots,T$, which imposes a rate constraint $\sum_{k=1}^K R_k \leq R_{R}$ where
\begin{multline}
R_{R}=\log_2\vert \mathbf{I}+\frac{1}{\sigma^2}(\mathbf{H}_{TR}+\mathbf{H}_{IR}\mathbf{\Theta}_1\mathbf{H}_{TI})\mathbf{W}\mathbf{W}^H\\
\times(\mathbf{H}_{TR}+\mathbf{H}_{IR}\mathbf{\Theta}_1\mathbf{H}_{TI})^H \vert,
\end{multline}
where $\mathbf{W}=[\mathbf{w_1}\ \mathbf{w_2},\ldots, \mathbf{w_k}]$. After decoding, the relay knows $s_k(t)$ for all $k=1,\ldots,K$ and $t=1,\ldots,T$, which can be assumed error free if $T$ is large enough and $\sum_{k=1}^K R_i\leq R_R$ is satisfied.

\subsection{Phase 2}
After decoding, the relay multiplies the decoded codewords by vectors $\mathbf{u}_k$ to beamform them to the RIS and the users. The relay sends $\sum_{k=1}^K\mathbf{u}_ks_k(t)$ leading to a transmit power $\frac{1}{2}\sum_{k=1}^K\|\mathbf{u}_k\|^2$.

The phase shifts at the RIS can be optimized independently from the ones used in first phase. Therefore, the received signal at the user $k$ in this phase becomes 
\begin{equation}
y_{k,2}(t)=(\mathbf{h}_{I,k}^H\mathbf{\Theta}_2\mathbf{H}_{IR}^H+\mathbf{h}_{R,k}^H)\sum_{i=1}^K\mathbf{u}_is_i(t)+n_2(t),
\end{equation}
$t=1,\ldots,T$, where $\mathbf{\Theta}_2=\text{diag}[e^{j\theta_{2,1}}, e^{j\theta_{2,2}},\ldots,e^{j\theta_{2,L}}]$ is the phase shift matrix in phase 2 with $\theta_{2,\ell}$ being the phase shift applied by the $\ell$th RIS element, and $n_2(t)$ is AWGN with zero mean and variance $\sigma^2$, independent through $t$. Note that we assume channel reciprocity between the relay and the RIS, and hence the relay-RIS channel is the Hermitian of the RIS-relay channel. 

User $k$ uses maximum ratio combining (MRC) to combine $y_{k,1}(t)$ and $y_{k,2}(t)$ before decoding $s_k(t)$, $t=1,\ldots,T$. Decoding is reliable if $R_k\leq\bar{R}_k$ where 
\begin{equation}
\bar{R}_{k}\leq \log_2(1+\gamma_{k,1}+\gamma_{k,2}),
\end{equation}
where $\gamma_{k,1}$ and $\gamma_{k,2}$ are the received signal-to-noise ratios (SNR) at user $k$ from the first and second phases, respectively, given by
\begin{align} 
\gamma_{k,1}&=\frac{\vert(\mathbf{h}_{I,k}^H\mathbf{\Theta}_1\mathbf{H}_{TI}+\mathbf{h}_{T,k}^H) \mathbf{w}_k\vert^2}{\sum_{i\neq k} \vert (\mathbf{h}_{I,k}^H\mathbf{\Theta}_1\mathbf{H}_{TI}+\mathbf{h}_{T,k}^H)\mathbf{w}_i\vert^2+\sigma^2}\\
\gamma_{k,2}&=\frac{\vert(\mathbf{h}_{I,k}^H\mathbf{\Theta}_2\mathbf{H}_{IR}^H+\mathbf{h}_{R,k}^H) \mathbf{u}_k\vert^2}{\sum_{i\neq k} \vert (\mathbf{h}_{I,k}^H\mathbf{\Theta}_2\mathbf{H}_{IR}^H+\mathbf{h}_{R,k}^H)\mathbf{u}_i\vert^2+\sigma^2}.
\end{align}

Next, we formulate the optimization problem for minimizing the total power subject to QoS constraints.

%

\section{Problem Formulation} \label{Sec:Form}
Our goal is to investigate the gains achieved by combining an RIS and a relay in terms of energy efficiency. Through this, we aim to study the impact of optimizing the beamforming matrices at the BS and the relay and the phase shifts at the RIS on the total transmit power under the QoS constraints, and to find which of the RIS and the relay contributes more towards minimizing the total power. To this end, we formulate the optimization problem as minimizing the power at the BS and the relay under QoS constraints, then we propose different solutions for the formulated problem. 

The quality of service constraint is given by a threshold rate $R_{th}$ so that each user can achieve this rate. This implies that the achievable rate $R_k/2$ has to be larger than $R_{th}$. The problem can thus be formulated as follows
\begin{subequations}
\label{OP1}
\begin{eqnarray}
&\displaystyle\min_{\mathbf{W}, \mathbf{\Theta}_1, \mathbf{\Theta}_2, \mathbf{U}}& \frac{1}{2}\sum_{k=1}^K\|\mathbf{w}_k\|^2+\frac{1}{2}\sum_{k=1}^K\|\mathbf{u}_k\|^2 \\
\label{OP1b}
&\text{s.t.}&   R_{R}\geq 2KR_{th},\\
\label{OP1c}
&& \bar{R}_{k}\geq 2R_{th},\ \ k=1,\ldots, K,\\
\label{OP1d}
&&\mathbf{\Theta}_1, \mathbf{\Theta}_2 \in \mathcal{F},
\end{eqnarray}
\end{subequations} 
where $\mathbf{U}=[\mathbf{u}_1,\ldots,\mathbf{u}_K]$, and $\mathcal{F}$ is the set of all possible discrete phase shifts. The objective function \eqref{OP1} is the total transmit power at the BS and the relay. Constraints \eqref{OP1b} and \eqref{OP1c} guarantees that the required QoS at users is achievable in both the first and second phases, respectively.

Note that all constraints of this problem are nonconvex due to the coupling between $\mathbf{W}$, $\mathbf{U}$, $\mathbf{\Theta}_1$, and $\mathbf{\Theta}_2$. Moreover, constraint \eqref{OP1d} restricts the phase shifts to be discrete. Even if the variables $\mathbf{\Theta}_1$ and $\mathbf{\Theta}_2$ are given, constraint \eqref{OP1c} is still nonconvex due to the coupling between $\mathbf{W}$ and $\mathbf{U}$. Hence, problem \eqref{OP1} is a mixed integer non-linear program (MINLP), which is NP-hard. There is no standard solution for such problem that finds the global optimum. However, in what follows, we propose two efficient solutions based on alternating optimization, singular value decomposition (SVD), and uplink-downlink duality. We first propose a solution for both $\mathbf{W}$ and $\mathbf{U}$ when $\mathbf{\Theta}_1$ and $\mathbf{\Theta}_2$ are given. Then we propose an outer loop to optimize both $\mathbf{\Theta}_1$ and~$\mathbf{\Theta}_2$.


\subsection{Optimizing $\mathbf{W}$ and $\mathbf{U}$ Given $\mathbf{\Theta}_1$ and $\mathbf{\Theta}_2$}
For given $\mathbf{\Theta}_1$ and $\mathbf{\Theta}_2$, problem \eqref{OP1} reduces to 
\begin{subequations}
\label{OP11}
\begin{eqnarray}
&\displaystyle\min_{\mathbf{W}, \mathbf{U}}& \frac{1}{2}\sum_{k=1}^K\|\mathbf{w}_k\|^2+\frac{1}{2}\sum_{k=1}^K\|\mathbf{u}_k\|^2 \\
\label{OP11b}
&\text{s.t.}&   R_{R}\geq 2KR_{th},\\
\label{OP11c}
&& \bar{R}_{k}\geq 2R_{th},\ \ k=1,\ldots, K,
\end{eqnarray}
\end{subequations} 
Constraints \eqref{OP11c} still makes the problem nonconvex. Hence, we propose a suboptimal, yet efficient, solutions as follows.

\subsubsection{SVD and Uplink-Downlink Duality} 
Due to the assumption that the relay and RIS are much closer to the users than the BS, the channel gains to the users in the first phase (from the BS) are weak compared to those in the second phase (from the relay). Hence, ignoring the channel gains to users in the first phase should cause negligible effect on the solution optimality. In other words, since optimizing $\mathbf{W}$ would produce a higher impact on $R_R$ than on $\bar{R}_k \ \forall k$, we design $\mathbf{W}$ to achieve constraint \eqref{OP11b} while ignoring \eqref{OP11c}.   
With that said, the problem can be separated into two optimization problems, one for $\mathbf{W}$ and the other for $\mathbf{U}$ with a given $\mathbf{W}$. This means that the matrix $\mathbf{W}$ can be found by solving the following problem

\begin{subequations}
\label{OP2}
\begin{eqnarray}
&\displaystyle\min_{\mathbf{W}}& \frac{1}{2}\sum_{k=1}^K\|\mathbf{w}_k\|^2 \\
\label{OP2b}
&\text{s.t.}&   R_{R}\geq 2KR_{th},
\end{eqnarray}
\end{subequations}  

The solution of \eqref{OP2} can be achieved using SVD with water filling. Specifically, by defining $\mathbf{H}_{TIR}=\mathbf{H}_{TR}+\mathbf{H}_{IR}\mathbf{\Theta}_1\mathbf{H}_{TI}=\bar{\mathbf{V}}\mathbf{\Lambda}^{\frac{1}{2}}\mathbf{V}^H$, where $\bar{\mathbf{V}}$ and $\mathbf{V}$ are the matrices of singular vectors and $\mathbf{\Lambda}^{\frac{1}{2}}$ is the matrix of singular values, the solution of $\mathbf{W}$ is given by 
\begin{equation}
\label{SVD}
\mathbf{W}=\mathbf{V}\bar{\mathbf{\Xi}}^{\frac{1}{2}},
\end{equation}
where $\bar{\mathbf{\Xi}}=\text{diag}[P_1, P_1,\ldots,P_K]$ is the power allocation matrix. Using \eqref{SVD}, problem \eqref{OP2} reduces to 
\begin{subequations}
\label{OP3}
\begin{eqnarray}
&\displaystyle\min_{\mathbf{P}}& \frac{1}{2}\sum_{k=1}^K P_k \\
\label{OP3b}
&\text{s.t.}&   \sum_{k=1}^K\log_2(1+\frac{1}{\sigma^2}P_k\lambda_k)\geq 2KR_{th},
\end{eqnarray}
\end{subequations}     
where $\lambda_k$ is the $k$th eigenvalue of the matrix $\mathbf{H}_{TIR}\mathbf{H}_{TIR}^H$ (square of diagonal components of $\Lambda^{\frac{1}{2}}$). Problem \eqref{OP3} is convex and can be solved using the water filling. The optimal $P_k$ is given by 
\begin{equation}
\label{Pk}
P_k=\left(\mu-\frac{\sigma^2}{\lambda_k}\right)^+,\ k=1,\ldots,K,
\end{equation} 
where $\mu$ is a dual variable and is given by $$\mu=\frac{\sigma^2e^{2R_{th}\log(2)}}{(\prod_{k=1}^K\lambda_k)^{\frac{1}{K}}}.$$  Therefore, $\mathbf{W}$ can be found using equations \eqref{Pk} $\forall k$ and \eqref{SVD}. 

Now, we can solve problem \eqref{OP1} for $\mathbf{U}$ using the obtained $\mathbf{W}$. The problem in terms of $\mathbf{U}$ can be expressed as follows
\begin{subequations}
\label{OP4}
\begin{eqnarray}
&\displaystyle\min_{\mathbf{U}}& \frac{1}{2}\sum_{k=1}^K\|\mathbf{u}_k\|^2 \\
\label{OP4c}
&\text{s.t.}& \frac{\vert(\mathbf{h}_{I,k}^H\mathbf{\Theta}_2\mathbf{H}_{RI}+\mathbf{h}_{R,k}^H) \mathbf{u}_k\vert^2}{\sum_{i\neq k} \vert (\mathbf{h}_{I,k}^H\mathbf{\Theta}_2\mathbf{H}_{RI}+\mathbf{h}_{R,k}^H)\mathbf{u}_i\vert^2+\sigma^2}\geq \eta_k,
\nonumber\\
&&\ \ \ \ \ \ \  k=1,\ldots, K,
\end{eqnarray}
\end{subequations} 
where   $\eta_k=2^{2R_{th}}-1-\frac{\vert(\mathbf{h}_{I,k}^H\mathbf{\Theta}_1\mathbf{H}_{TI}+\mathbf{h}_{T,k}^H) \mathbf{w}_k\vert^2}{\sum_{i\neq k} \vert (\mathbf{h}_{I,k}^H\mathbf{\Theta}_1\mathbf{H}_{TI}+\mathbf{h}_{T,k}^H)\mathbf{w}_i\vert^2+\sigma^2}$.
The optimal solution of problem \eqref{OP4} can be found using either semidefinite programming, second order cone programming, or the uplink-downlink duality. For the sake of simplicity, we adopt the uplink-downlink duality approach. The optimal solution of problem \eqref{OP4} can be found using
\begin{equation}
\label{u}
\mathbf{u}_k=\sqrt{q_k}\bar{\mathbf{u}}_k, k=1,\ldots,K,
\end{equation}
where $q_k$ is the $k$th entry of vector  
\begin{equation}
\label{q}
\mathbf{q}=\sigma^2\mathbf{D}^{-1}\mathbf{1}_{K},
\end{equation}
$\mathbf{1}_{K}$ is the one vector with length $K$, $\mathbf{D}$ and $\bar{\mathbf{u}}_k$ are given by
\begin{equation}
\mathbf{D}_{i,j}=\begin{cases} \frac{\vert \mathbf{h}_{RI,i} \bar{\mathbf{u}}_i\vert^2}{\eta_i}, & i=j;\\
-\vert\mathbf{h}_{RI,i} \bar{\mathbf{u}}_j\vert^2, & i\neq j,
\end{cases}
\end{equation}
\begin{equation}
\label{ubar}
\bar{\mathbf{u}}_k=\frac{(\mathbf{I}_N+\sum_{i=1}^K \frac{\beta_i}{\sigma^2}\mathbf{h}_{RI,i}\mathbf{h}_{RI,i}^H)^{-1}\mathbf{h}_{RI,k}}{\|(\mathbf{I}_N+\sum_{i=1}^K \frac{\beta_i}{\sigma^2}\mathbf{h}_{RI,i}\mathbf{h}_{RI,i}^H)^{-1}\mathbf{h}_{RI,k}\|}, \forall k,
\end{equation}
where $\mathbf{h}_{RI,i}=\mathbf{h}_{I,i}^H\mathbf{\Theta}_2\mathbf{H}_{RI}+\mathbf{h}_{R,i}^H$, and 
\begin{equation}
\label{beta}
\beta_k=\frac{\sigma^2}{(1+\frac{1}{\eta_k})\mathbf{h}_{RI,k}^H(\mathbf{I}_N+\sum_{i=1}^K \frac{\beta_i}{\sigma^2}\mathbf{h}_{RI,i}\mathbf{h}_{RI,i}^H)^{-1}\mathbf{h}_{RI,k}}.
\end{equation}
Thus, first we find the dual variable $\beta_k \ \forall k$ using the fixed point algorithm on \eqref{beta}. Then, $\mathbf{u}_k$ can be found using equations \eqref{ubar}-\eqref{u}. 

\subsubsection{SVD and Zero-Forcing}
Since the above method does not provide a closed-form solution for $\mathbf{U}$ and includes an iterative procedure, we propose another simpler solution next. Simple solutions for $\mathbf{W}$ and $\mathbf{U}$ are desirable because they would be implemented several times to find a good solution for $\mathbf{\Theta}_1$ and $\mathbf{\Theta}_2$. Therefore, we propose to design $\mathbf{U}$ to eliminate the interference at the users. This can be obtained using zero-forcing (ZF), where $\mathbf{U}$ is given by

\begin{equation}
\label{UZF}
\mathbf{U}=\mathbf{H}_{RIK}^H(\mathbf{H}_{RIK}\mathbf{H}_{RIK}^H)^{-1}\mathbf{Q}^{\frac{1}{2}}.
\end{equation}
where the $k$th column of $\mathbf{H}_{RIK}$ is $\mathbf{h}_{RI,k}$ and $\mathbf{Q}=\text{diag}[q_1,q_2,\ldots,q_K]$ is the power allocation matrix at the relay. With this design and assumption, problem \eqref{OP1} becomes
\begin{subequations}
\label{OP5}
\begin{eqnarray}
&\displaystyle\min_{\mathbf{q}}&   \frac{1}{2}\text{tr}(\mathbf{U}\mathbf{U}^H) \\
\label{OP5c}
&\text{s.t}& q_k\geq \sigma^2\eta_k,\ \ k=1,\ldots, K,
\end{eqnarray}
\end{subequations} 
the solution of which is given by $q_k=\sigma^2\eta_k$. Note that in this method, $\mathbf{W}$ is chose as in \eqref{SVD}, i.e., using the SVD method.

\subsection{Optimizing $\mathbf{\Theta}_1$ and $\mathbf{\Theta}_2$}

After finding solutions for $\mathbf{W}$ and $\mathbf{U}$, we find solutions for $\mathbf{\Theta}_1$ and $\mathbf{\Theta}_2$. Finding $\mathbf{\Theta}_1$ and $\mathbf{\Theta}_2$ can be expressed as follows 
\begin{subequations}
\label{OP6}
\begin{eqnarray}
&\displaystyle\min_{\mathbf{\Theta}_1, \mathbf{\Theta}_2}&  \frac{1}{2}\text{tr}(\mathbf{V}\mathbf{\Xi}\mathbf{V}^H) + \frac{1}{2}\text{tr}(\mathbf{U}\mathbf{U}^H) \\
\label{OP6b}
&\text{s.t.}&   R_{R}\geq 2KR_{th},\\
\label{OP6c}
&& \bar{R}_k\geq 2R_{th},\ \ k=1,\ldots, K,\\
\label{OP6d}
&&\mathbf{\Theta}_1, \mathbf{\Theta}_2 \in \mathcal{F},
\end{eqnarray}
\end{subequations} 
Providing an optimal solution for the above problem is difficult since relating $\mathbf{\Theta}_1$ and $\mathbf{V}$ is not straightforward, $\mathbf{\Theta}_2$ is involved in calculating inverse matrices to find $\mathbf{U}$, both $\mathbf{\Theta}_1$ and $\mathbf{\Theta}_2$ are coupled through \eqref{OP6c}, and both $\mathbf{\Theta}_1$ and $\mathbf{\Theta}_2$ are discrete. Hence, proposing a heuristic solution is appropriate for such a problem. Authors of \cite{cui2014coding, kaina2014shaping} show that the practical number of bits of each phase shift is limited (e.g., $b=1$ or $b=2$). This means that the search space of each phase shift is small if the other phase shifts are fixed. Therefore, we adopt the approach that fixes $2L-r$ phase shifts and finds the optimal solution for the remaining $r$ phase shifts using an $r$-dimensional search. We repeat this until we go over all the phase shifts in $\mathbf{\Theta}_1$ and $\mathbf{\Theta}_2$ several rounds. 
Increasing $r$ would increase complexity but improve performance.

\subsection{Baseline Approaches}
\label{BA}
We adopt the following approaches and models for comparison purposes.

\subsubsection{Relay-Assisted Multi-user System} 
We solve the optimization problem \eqref{OP1} when the RIS is absent to evaluate the impact of the relay only. The problem in this case is solved using the same procedures, but $\mathbf{\Theta}_1$ and $\mathbf{\Theta}_2$ are zero. The goal of this approach is to quantify the amount of power reduction achieved using the RIS.

\subsubsection{RIS-Assisted Multi-user System}
We also solve the problem when the relay is absent and the system consists of BS, RIS, and multiple users. In this case, the half-duplex constraint is dropped and the achievable rates are multiplied by two. To guarantee a fair comparison, the transmit power at the BS is multiplied by two because the BS transmit power in this case is double the power of half-duplex case. The problem can then be solved without considering $\mathbf{U}$ and $\mathbf{\Theta}_2$. The problem of optimizing $\mathbf{W}$ becomes the same as problem \eqref{OP4} and can be solved using uplink-downlink duality, while $\mathbf{\Theta}_1$ can be found by fixing $L-r$ phases and using multiple rounds of $r$-dimensional searches.        

\section{Simulation Results}\label{Sec:Sim}
\begin{figure}[!t]
\centering
\includegraphics[width=3in]{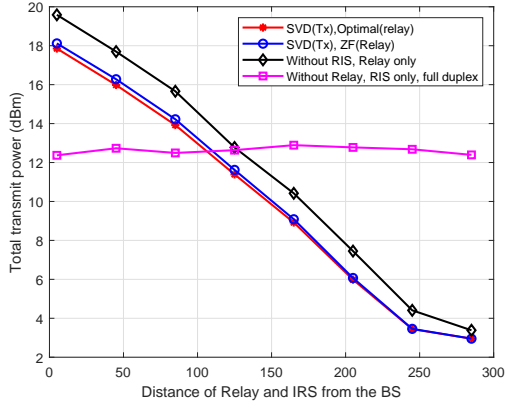}
\caption{The required total power versus the distance from relay and RIS to the BS, where distance between BS and the center of users' circle is $300$ m, $K=7$, $N=9$, $M=10$, $L=50$, and $R_{th}=2$.}
\label{P_Dis}
\end{figure}

In the following simulations, we assume that the antennas at the BS and relay form a uniform linear array (ULA), while the reflecting elements at the RIS form a uniform planner array (UPA). The distances between the antennas and reflecting elements are calculated based on this deployment. The users are randomly distributed inside a circle of radius $\rho=40$ m, whose center is $300$ m away from the BS.  The relay and the RIS are assumed to be half-way between the BS and the users' circle center (except in Fig. \ref{P_Dis}). In all simulations, unless otherwise noted, we assume that $M=10$, $N=9$, and $L=50$. The number of bits $b$ is assumed to be $b=2$, so each reflecting element can have $2^b$ different phases, which are $\mathcal{F}=[0, \Delta \theta, \ldots, (2^b-1)\Delta \theta]$, where $\Delta \theta=\frac{2 \pi}{2^b}$. 
\begin{figure}[!t]
\centering
\includegraphics[width=3in]{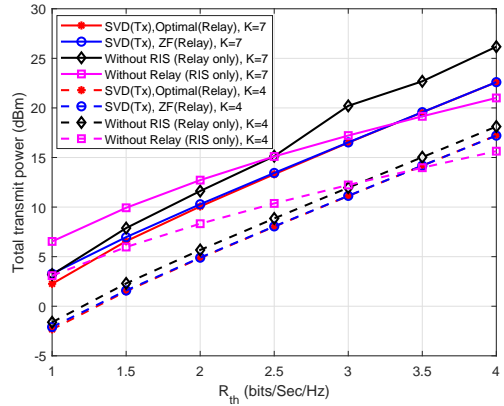}
\caption{The required total power versus the required $R_{th}$ with different number of users $K$ when $N=9$, $M=10$, $L=50$.}
\label{P_Rth}
\end{figure}

We evaluate the impact of changing the required rates at the users ($R_{th}$), the number of users, and the location of both the relay and the RIS on the required total power of the system. In all figures, we compare the proposed system and the baselines discussed in Sec. \ref{BA}. We assume that the channels between BS-relay, BS-RIS, and relay-RIS are all modelled as a Rician fading channel model, where the LoS is available. Whereas, the channels between any point to the any user is assumed to be Rayleigh fading channel, where the LoS is not available. The channel attenuation coefficient between any two points is given by $\beta=\frac{C}{d^{\alpha}}$, where $C=10^{\frac{G_t+G_r-35.95}{10}}$, $\alpha=2.2$ if the LoS is available, and $C=10^{\frac{G_t+G_r-33.95}{10}}$, $\alpha=3.67$ if the LoS is not available, where $G_t=5$ dBi and $G_r=0$ dBi are the antenna gains in dBi at the transmitter and the receiver \cite{NadeemR4}.       
\begin{figure}[!t]
\centering
\includegraphics[width=3in]{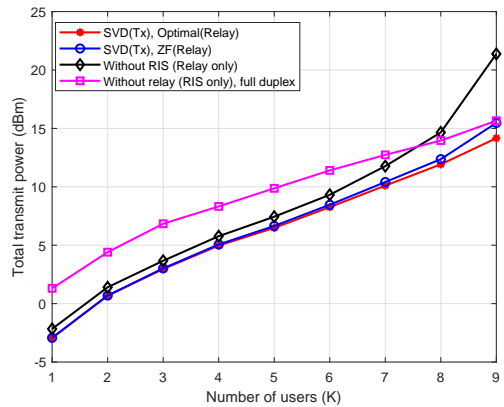}
\caption{The required total power versus the number of users in the system when $N=9$, $M=10$, $L=50$, and $R_{th}=2$.}
\label{P_K}
\end{figure}

Fig. \ref{P_Rth} shows how the total power behaves as a function of $R_{th}$ with the different approaches and different number of users. The figure shows that the  coexistence of relay and RIS improves the energy efficiency significantly especially when the required $R_{th}$ is not large. When the required $R_{th}$ increases beyond a certain value, the system with only RIS (i.e., without relay) starts to perform better than the system with both a relay and an RIS. This is due to the assumed half duplex operation when the relay is present, and this conclusion would change if the system is operated in a full-duplex mode. The figure also shows that the two proposed approaches (i.e., SVD-optimal beamforming and SVD-ZF) almost perform the same except when $R_{th}$ is small. Fig. \ref{P_Rth} also shows that increasing the number of users leads to increasing the required power to achieve the same required $R_{th}$, and at the same time the performance gap between the proposed system and the system without RIS increases. To make this point clearer, we plot the number of users versus the required power in Fig. \ref{P_K}.

Fig. \ref{P_K} shows that increasing the number of users in the system leads to increasing the required power to achieve an $R_{th}=2$. It can be seen that as $K$ increases the total required power at the system increases, especially when the system consists only of relay (i.e., without RIS).  In other words, the number of antennas at the relay should be much higher than the number of users to improve the energy efficiency of the system without RIS (relay only). The figure also shows that the optimal approach at the relay (uplink-downlink duality approach) performs better than the ZF when $K$ get closer to $N$. The figure shows that both relay and RIS improves the energy efficiency of the system even when the system works under the half-duplex constraint. 

Fig. \ref{P_Dis} shows how the required power behaves if we change the location of both the relay and the RIS. We assume that the distance from the BS to the center of the user' circle is fixed and both RIS and the relay change their distances by moving from near the BS to near the users' center. The figure shows that the RIS contribution to the system performance is not significantly affected by the distance, but the relay's contribution is. Since a line-of-sight path is assumed to exist between the BS, RIS, and relay, the second hop (between relay, RIS and users) is the weaker link. This explains why the system performance improves as the relay and RIS move toward the users, since the channel quality of the second hop improves. The figure indicates that, the best location of both relay and RIS is in the center of the users as long as a LoS exists in the first hop and is absent in the second one. 

\section{Conclusion}\label{Sec:Conc}
This paper proposes the coexistence of relay and RIS to improve the energy efficiency of multiuser systems. The paper also proposes solutions for joint beamforming at the BS, the relay, and the RIS to minimize the required total power (at the BS and the relay) under given QoS at the users. Since the formulated optimization is difficult, we propose suboptimal solution the relies on the fact that the equivalent channel from the BS to the users is weak relative to the relay-user channel.  In the simulation results, we evaluate and compare the contribution of both the relay and the RIS under changing different system parameters. In general, the proposed system improves the energy efficiency significantly, especially when the required $R_{th}$ is not large and when the relay and RIS are closer to the users than to the BS.


\bibliographystyle{IEEEtran}

\bibliography{mylib}

%
%
%
%
%
%
%
%

\end{document}